\documentclass[final,5p,times,twocolumn]{elsarticle}

\usepackage[utf8]{inputenc}
\usepackage{amsmath,amssymb,amsfonts}
\usepackage{bm}
\usepackage{graphicx}
\usepackage{xcolor}
\usepackage{booktabs}
\usepackage[colorlinks=true,breaklinks=true,linkcolor=blue,citecolor=purple,urlcolor=teal]{hyperref}
\usepackage{url}

\renewcommand{\Im}{\mathop{\mathrm{Im}}}

\journal{Physics Letters B}

\biboptions{comma,sort&compress}

\begin{document}

\begin{frontmatter}

\title{Coulomb bridge mechanism for peripheral polarization of weakly bound projectiles}

\author[tongji]{Hao Liu}
\author[tongji,scnt]{Jin Lei\corref{cor1}}
\cortext[cor1]{Corresponding author}
\ead{jinl@tongji.edu.cn}
\author[tongji]{Zhongzhou Ren}
\address[tongji]{School of Physics Science and Engineering, Tongji University, Shanghai 200092, China}
\address[scnt]{Southern Center for Nuclear-Science Theory (SCNT), Institute of Modern Physics, Chinese Academy of Sciences, Huizhou 516000, Guangdong Province, China}

\begin{abstract}
Peripheral reactions of weakly bound nuclei probe how continuum excitation feeds back on the elastic channel. For compact weakly bound projectiles this feedback is expected to be concentrated near the nuclear surface, whereas for halo projectiles the dilute projectile tail and strong low-energy dipole response can let the long-range Coulomb field carry polarization at impact parameters where nuclear absorption is already weak. We formulate this physical separation through the Feshbach dynamical polarization potential (DPP) in the continuum-discretized coupled-channels (CDCC) framework. The DPP has a bridge-propagator-bridge structure: the elastic ($P$) channel is connected to the reaction ($Q$) subspace by the two $P\!\leftrightarrow\!Q$ couplings $U_{0\gamma}$ and $U_{\gamma'0}$, linked by the coupled-channel propagator $g_{\gamma\gamma'}$ inside $Q$. Splitting these bridge couplings into nuclear and Coulomb parts, while retaining one common $Q$-space propagator, gives $\Delta U_{\rm DPP}=\Delta U_N+\Delta U_C+\Delta U_{NC}$. Applied to $d+{}^{58}\mathrm{Ni}$, ${}^{6}\mathrm{Li}+{}^{208}\mathrm{Pb}$, ${}^{11}\mathrm{Be}+{}^{64}\mathrm{Zn}$, and ${}^{8}\mathrm{B}+{}^{64}\mathrm{Zn}$, the decomposition shows a clear system dependence: a nuclear bridge in the light system, a mixed bridge with strong destructive interference in the heavy stable-projectile system, and a Coulomb-dominated bridge in both halo cases, with the proton-halo system showing the most pronounced Coulomb sector and a constructive nuclear--Coulomb interference. For the halo reactions, peripheral partial waves ($L\gtrsim35$) satisfy $\sigma_R^L\simeq\sigma_{\rm DPP}^L\simeq\sigma_{\rm BU}^L$, and the high-$L$ DPP tail is dominated by $\sigma_C^L$. Removing off-diagonal Coulomb propagation inside $Q$ leaves this behavior largely intact, whereas removing the $P\!\leftrightarrow\!Q$ Coulomb bridge collapses both DPP-induced absorption and breakup. The peripheral polarization of the halo reactions is therefore identified as a Coulomb-bridge effect, and the high-$L$ elastic-breakup yield serves as an observable signature of the same Coulomb $P\!\leftrightarrow\!Q$ bridge couplings.
\end{abstract}


\end{frontmatter}

\section{Introduction}
\label{sec:intro}

Near the Coulomb barrier, weakly bound projectiles such as ${}^{6,8}\mathrm{He}$, ${}^{6,7}\mathrm{Li}$, ${}^{11}\mathrm{Be}$, and ${}^{8}\mathrm{B}$ can be polarized and broken up before the projectile reaches the strongly absorptive nuclear surface~\cite{Tanihata1985,Hansen1995,Jonson2004,Canto2006,Canto2015,Tanihata2013}. The elastic channel then remembers the open continuum: part of the incident flux is lost into breakup and fragment absorption, and the surviving elastic wave is refracted by virtual excursions through those channels. This feedback is described by the nonlocal, energy-dependent dynamical polarization potential (DPP), whose imaginary part measures continuum-induced absorption and whose real part modifies the effective refraction~\cite{Mackintosh1982,Sakuragi1987,Thompson1989}. The continuum-discretized coupled-channels (CDCC) method~\cite{Rawitscher1974,Austern1987,Johnson1989,Yahiro2012,ThompsonNunes,Matsumoto2004,RodriguezGallardo2008}, combined with the Feshbach projection formalism~\cite{Feshbach1958,Feshbach1962,Keeley2004,Mackintosh2012}, provides the standard framework for following this feedback.

\begin{figure}[t]
\centering
\includegraphics[width=\columnwidth]{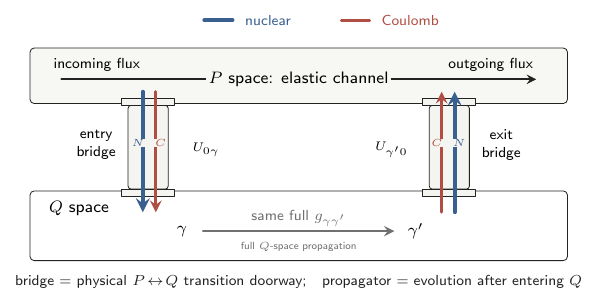}
\caption{Bridge decomposition of the Feshbach dynamical polarization potential. The elastic channel $P$ is connected to the reaction subspace $Q$ through two $P\!\leftrightarrow\!Q$ bridge couplings, $U_{0\gamma}$ and $U_{\gamma'0}$, each of which admits a natural splitting into nuclear and Coulomb components. The coupled-channel propagator $g_{\gamma\gamma'}$ inside $Q$ is the same complete solution for every term of the decomposition. The nuclear--Coulomb separation is therefore made at the doorway couplings, not by assigning separate $Q$-space propagators to nuclear and Coulomb reaction paths.}
\label{fig:bridge}
\end{figure}

Partitioning the Hilbert space into an elastic subspace $P$ and its complement $Q$, the DPP reads
\begin{equation}
\Delta U_{\rm DPP}(R,R') = \sum_{\gamma,\gamma'\in Q} U_{0\gamma}(R)\,g_{\gamma\gamma'}(R,R')\,U_{\gamma'0}(R'),
\label{eq:dpp_general}
\end{equation}
where $g_{\gamma\gamma'}$ is the coupled-channel Green's function in $Q$~\cite{Liu2026Green} and $U_{0\gamma}$, $U_{\gamma'0}$ are the $P\!\leftrightarrow\!Q$ bridge couplings. This form gives a simple physical picture~\cite{Liu2025Feshbach,Liu2026Coherent}: flux leaves the elastic channel through one bridge coupling, propagates through the breakup space, and returns or is absorbed through the second. In a series of earlier papers we formulated the full-coupling Feshbach DPP within CDCC, established the existence and uniqueness of the coupled-channel Green's function entering Eq.~(\ref{eq:dpp_general}), partitioned the absorption cross section spatially through an ingoing-wave boundary condition, and decomposed the total non-elastic cross section into elastic-breakup, fragment-absorption, and interference contributions via a generalized optical theorem~\cite{Liu2025Feshbach,Liu2026Green,Liu2026PLB,Liu2026Coherent}. The present Letter goes one step further by splitting the $P\!\leftrightarrow\!Q$ bridge couplings themselves into nuclear and Coulomb parts, so that the matrix elements responsible for peripheral polarization can be identified individually. The question addressed here is therefore not only whether Coulomb couplings matter, but whether the peripheral polarization is carried mainly by the short-range nuclear bridge or by the long-range Coulomb bridge. Figure~\ref{fig:bridge} summarizes this organizing picture.

For weakly bound projectiles, two physically distinct mechanisms populate this bridge. The short-range nuclear interaction drives transitions through the gradient of the fragment--target optical potentials at the nuclear surface, while the long-range Coulomb field generates strong electric multipole transitions that dominate at low energies on heavy targets~\cite{Alder1956,Baur1986,Bertulani1988}. Their balance depends sensitively on the target charge and the incident energy, and their interference is decisive around the Coulomb barrier~\cite{CantoGomes2009,DiPietro2010,DiPietro2012}. Earlier studies have addressed this competition with switch-off calculations in CDCC or perturbative schemes, by removing either the nuclear or the Coulomb coupling and comparing the result with the full calculation~\cite{Nunes1999,Tostevin2002,Summers2007,Moro2009,Crespo2007,Rusek2011,Moro2019}. Such calculations answer a sensitivity question and have shown that the two contributions interfere and cannot simply be added. They do not, however, identify the responsible matrix elements, because switching off a coupling also alters the coupled-channel propagator. A related decomposition at the scattering-matrix level was carried out by P.~Descouvemont \emph{et al.}~\cite{pierre17}, who split the CDCC breakup amplitude into nuclear and Coulomb components through a Lippmann--Schwinger reduction for ${}^{7}\mathrm{Li}+{}^{208}\mathrm{Pb}$; they found that at peripheral partial waves the nuclear contribution becomes negligible and the Coulomb part is well defined, while the splitting at small $L$ remains method-dependent because both components share the same coupled-channel wave function. Isolating the dynamical role of each sector of the bridge requires a decomposition that splits $U_{0\gamma}$ and $U_{\gamma'0}$ alone, while keeping a single $Q$-space propagator $g_{\gamma\gamma'}$ common to every component.

This Letter follows that physical picture at the operator level. Splitting each bridge coupling into its nuclear and Coulomb parts and substituting into Eq.~(\ref{eq:dpp_general}) gives three components, $\Delta U_N$, $\Delta U_C$, and $\Delta U_{NC}$, that share the same $Q$-space propagator and quantify which sector of the bridge transports flux from the elastic channel. The applications are arranged as a controlled hierarchy. The reaction $d+{}^{58}\mathrm{Ni}$ provides a nuclear-bridge baseline, ${}^{6}\mathrm{Li}+{}^{208}\mathrm{Pb}$ exposes a mixed nuclear--Coulomb bridge with strong destructive interference, and the halo systems ${}^{11}\mathrm{Be}+{}^{64}\mathrm{Zn}$ and ${}^{8}\mathrm{B}+{}^{64}\mathrm{Zn}$ reach the limit in which peripheral polarization is carried predominantly by $U^{(C)}_{0\gamma}$ and $U^{(C)}_{\gamma'0}$.

\section{Nuclear--Coulomb decomposition of the DPP}
\label{sec:theory}

In practical CDCC calculations the coupling form factors $U_{0\gamma}$ originate from the multipole expansion of the fragment--target interactions and admit the natural separation
\begin{equation}
\begin{aligned}
U_{0\gamma}(R)   &= U^{(N)}_{0\gamma}(R)   + U^{(C)}_{0\gamma}(R),\\
U_{\gamma'0}(R') &= U^{(N)}_{\gamma'0}(R') + U^{(C)}_{\gamma'0}(R'),
\end{aligned}
\label{eq:coupling_split}
\end{equation}
into short-range nuclear and long-range Coulomb components. Substituting Eq.~(\ref{eq:coupling_split}) into Eq.~(\ref{eq:dpp_general}) yields
\begin{equation}
\begin{split}
\Delta U_{\rm DPP}(R,R') = {} & \Delta U_N(R,R') + \Delta U_C(R,R') \\
                              & {} + \Delta U_{NC}(R,R'),
\end{split}
\label{eq:dpp_decomp}
\end{equation}
with
\begin{subequations}
\label{eq:dpp_components}
\begin{align}
\Delta U_N(R,R')    &= \sum_{\gamma,\gamma'\in Q} U^{(N)}_{0\gamma}(R)\,g_{\gamma\gamma'}(R,R')\,U^{(N)}_{\gamma'0}(R'), \label{eq:dpp_N}\\
\Delta U_C(R,R')    &= \sum_{\gamma,\gamma'\in Q} U^{(C)}_{0\gamma}(R)\,g_{\gamma\gamma'}(R,R')\,U^{(C)}_{\gamma'0}(R'), \label{eq:dpp_C}\\
\Delta U_{NC}(R,R') &= \sum_{\gamma,\gamma'\in Q} \bigl[\, U^{(N)}_{0\gamma}(R)\,g_{\gamma\gamma'}(R,R')\,U^{(C)}_{\gamma'0}(R') \notag \\
                    &\quad + U^{(C)}_{0\gamma}(R)\,g_{\gamma\gamma'}(R,R')\,U^{(N)}_{\gamma'0}(R') \bigr]. \label{eq:dpp_NC}
\end{align}
\end{subequations}

Equation~(\ref{eq:dpp_decomp}) makes explicit that the DPP is not a simple sum of nuclear and Coulomb contributions taken independently: the two mechanisms are linked through the common $Q$-space propagator, which generates a finite interference term $\Delta U_{NC}$. The decomposition therefore isolates which part of the $P\!\leftrightarrow\!Q$ bridge feeds the polarization, while leaving the propagation inside $Q$ unchanged (see Fig.~\ref{fig:bridge}).

The DPP-induced loss of flux from the elastic channel,
\begin{equation}
\sigma_{\rm DPP} = -\frac{2}{\hbar v}\,\Im\langle\phi_0|\Delta U_{\rm DPP}|\phi_0\rangle,
\label{eq:sigma_dpp_def}
\end{equation}
where $\phi_0$ is the elastic-channel component of the full CDCC wave function~\cite{Liu2025Feshbach,Liu2026Coherent}, inherits the splitting
\begin{equation}
\sigma_{\rm DPP} = \sigma_N + \sigma_C + \sigma_{NC},
\label{eq:sigma_dpp}
\end{equation}
and likewise the partial-wave version $\sigma_{\rm DPP}^L = \sigma_N^L + \sigma_C^L + \sigma_{NC}^L$. The bare elastic-channel optical potential $U_{00}$, namely the diagonal element of the coupled-channel matrix before the DPP is added, produces a bare-potential absorption $\sigma_{U_{00}} = -(2/\hbar v)\,\Im\langle\phi_0|U_{00}|\phi_0\rangle$. Together with $\sigma_{\rm DPP}$ this saturates the total reaction cross section, $\sigma_R = \sigma_{U_{00}} + \sigma_{\rm DPP}$. Throughout this work the spin degrees of freedom of the projectile fragments are neglected to keep the coupled-channel problem tractable.

\section{Results}
\label{sec:results}

\subsection{A nuclear-bridge control case: $d+{}^{58}\mathrm{Ni}$}

For $d+{}^{58}\mathrm{Ni}$, the deuteron is treated as a $p$--$n$ system. The continuum is discretized for $\ell\le 2$. Proton and neutron optical potentials on ${}^{58}\mathrm{Ni}$ are taken from the Koning--Delaroche global parametrization (KD02)~\cite{KD02}. Calculations are performed at $E_d=20$, $25$, $28.6$, $30$, $35$, $40$, $52$, and $80$~MeV.

\begin{figure}[t]
\centering
\includegraphics[width=\columnwidth]{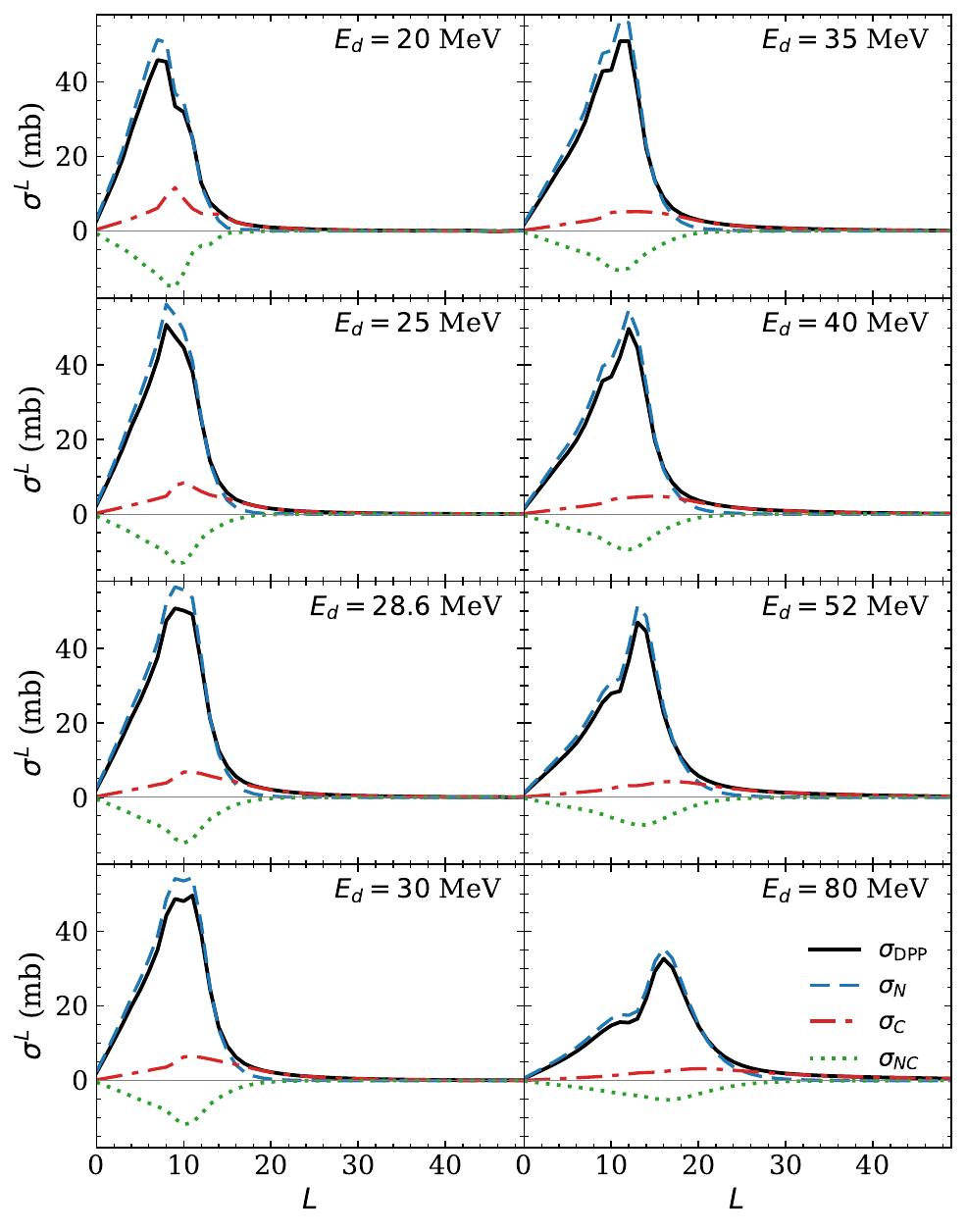}
\caption{Partial-wave decomposition of the DPP-induced absorption cross section for $d+{}^{58}\mathrm{Ni}$ at eight laboratory energies ($E_d=20$--$80$~MeV). Each panel shows the total $\sigma_{\rm DPP}^L$ (solid black), nuclear $\sigma_N^L$ (blue dashed), Coulomb $\sigma_C^L$ (red dash-dotted), and interference $\sigma_{NC}^L$ (green dotted).}
\label{fig:dNi}
\end{figure}

Figure~\ref{fig:dNi} shows the partial-wave decomposition of $\sigma_{\rm DPP}^L$ for $d+{}^{58}\mathrm{Ni}$ at the eight energies considered. At every energy, the nuclear component $\sigma_N^L$ (blue dashed) dominates the total $\sigma_{\rm DPP}^L$ (solid black) and closely tracks it in both shape and peak position. The Coulomb component $\sigma_C^L$ (red dash-dotted) is a broad distribution whose center of weight lies systematically at higher angular momentum than the nuclear one. This offset in the partial-wave distribution reflects the long-range, peripheral nature of the Coulomb coupling, which extends to partial waves beyond the grazing orbit, where the short-range nuclear coupling is already negligible. The interference $\sigma_{NC}^L$ (green dotted) is negative at essentially all $L$ and pointwise cancels a large fraction of the Coulomb contribution.

This pattern survives integration over angular momentum. The nuclear component $\sigma_N$ tracks the total $\sigma_{\rm DPP}$ across the full energy range, while the Coulomb component $\sigma_C$ remains relatively small in this deuteron-induced system and is largely compensated by the destructive interference contribution $\sigma_{NC}$. As a result, $\sigma_{\rm DPP}\approx\sigma_N$ to within a few percent. The first system therefore serves as the control case: when the nuclear bridge is strong and compact, the Coulomb sector is visible in the decomposition but gives only a small net correction after interference is included.

\subsection{A mixed bridge with destructive interference: ${}^{6}\mathrm{Li}+{}^{208}\mathrm{Pb}$}

For ${}^{6}\mathrm{Li}+{}^{208}\mathrm{Pb}$, ${}^{6}\mathrm{Li}$ is described as an $\alpha$+$d$ cluster, with the continuum states discretized into bins up to an $\alpha$-$d$ relative kinetic energy of $12$~MeV, with six bin states per partial wave and relative orbital angular momentum $\ell\le 2$. Calculations are performed at $E_{\rm lab}=29$, $33$, $39$, and $46$~MeV. The fragment--target optical potentials are taken from Refs.~\cite{alpha2,yyq06}; the surface imaginary part of the deuteron--target potential is removed to avoid double counting with the explicitly treated breakup channels, following Refs.~\cite{jin15-removingsur,jin17-removingsur}.

\begin{figure}[t]
\centering
\includegraphics[width=\columnwidth]{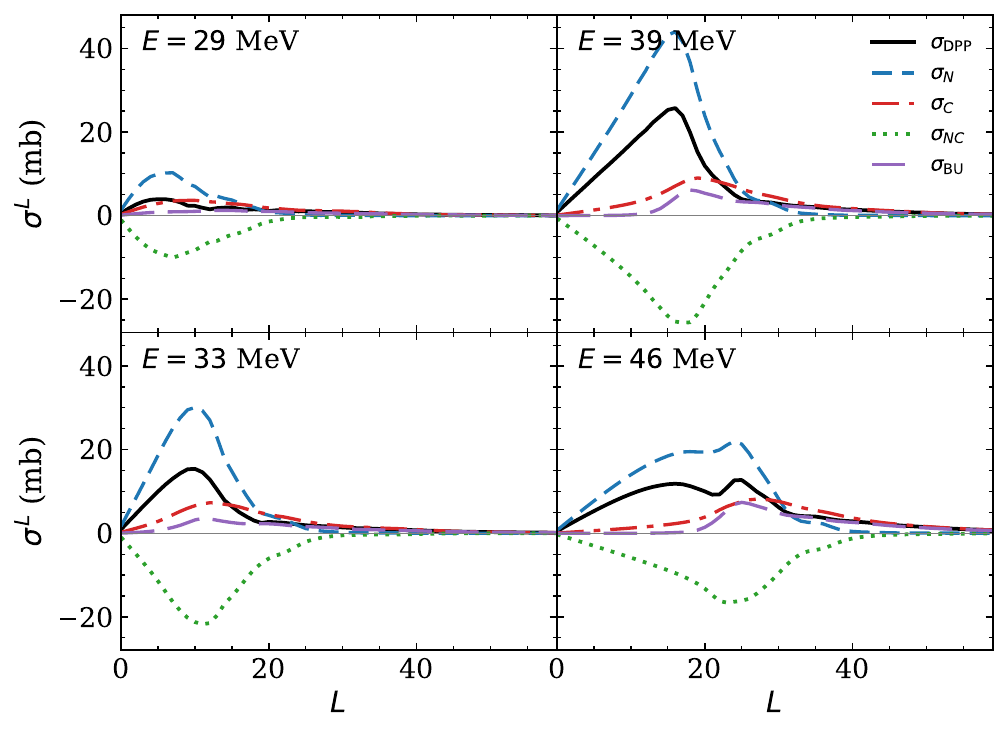}
\caption{Partial-wave decomposition of the DPP-induced absorption cross section for ${}^{6}\mathrm{Li}+{}^{208}\mathrm{Pb}$ at four laboratory energies. The panels show the total $\sigma_{\rm DPP}^L$ (solid black), nuclear $\sigma_N^L$ (blue dashed), Coulomb $\sigma_C^L$ (red dash-dotted), interference $\sigma_{NC}^L$ (green dotted), and elastic-breakup $\sigma_{\rm BU}^L$ (purple dash-dotted).}
\label{fig:LiPb}
\end{figure}

For this heavy-target case (Fig.~\ref{fig:LiPb}), the qualitative structure is similar, with a nuclear peak, a broader Coulomb distribution, and a negative interference, but the relative weights are very different. The partial cancellation between $\sigma_C^L$ and $\sigma_{NC}^L$ persists, while $\sigma_C^L$ now accounts for a significant fraction of the total $\sigma_{\rm DPP}^L$ and increases with the incident energy. At larger partial waves, the Coulomb component becomes comparable to, or even larger than, the nuclear component. Thus, ${}^{6}\mathrm{Li}+{}^{208}\mathrm{Pb}$ is the interpolation case in the hierarchy: the DPP-induced absorption is no longer purely nuclear dominated, but the Coulomb sector still appears through a coherent mixture with a large destructive interference term. ${}^{6}\mathrm{Li}$ is weakly bound but its $\alpha+d$ cluster structure, with two charged fragments and no soft-dipole strength close to threshold, differs qualitatively from the long-range single-nucleon tails of the halo systems considered below; it therefore sits at the compact-weakly-bound end of the hierarchy rather than as a halo case.

In Fig.~\ref{fig:LiPb}, the elastic-breakup cross section $\sigma_{\rm BU}^L$ is also shown by the purple dashed line. It follows the broad high-$L$ structure of $\sigma_{\rm DPP}^L$ at every energy, indicating that the DPP-induced flux loss and the asymptotic breakup yield are already correlated in the peripheral region. The correspondence is not yet one-to-one, since $\sigma_{\rm DPP}$ also contains continuum-induced absorption and coherent interference terms~\cite{Liu2026Coherent}. The important point for the present argument is more specific: in the high-$L$ region the breakup profile begins to track the Coulomb component of the DPP decomposition. This motivates the halo test below: the limiting relation $\sigma_{\rm BU}^L\simeq\sigma_{\rm DPP}^L\simeq\sigma_C^L$ is expected to emerge cleanly only when a small valence separation energy and a strong low-lying dipole response push the breakup form factor close to threshold, so that the Coulomb bridge dominates the $P\!\leftrightarrow\!Q$ coupling even on a target lighter than ${}^{208}\mathrm{Pb}$. The neutron-halo system ${}^{11}\mathrm{Be}+{}^{64}\mathrm{Zn}$ ($Z_T=30$, $S_n=504$~keV) sits precisely in this regime.

\subsection{The halo limit: Coulomb bridge controls peripheral polarization}

For ${}^{11}\mathrm{Be}+{}^{64}\mathrm{Zn}$ at $E_{\rm lab}=28.7$~MeV, ${}^{11}\mathrm{Be}$ is modeled as a ${}^{10}\mathrm{Be}+n$ two-body system. The continuum is discretized up to $\varepsilon_{\max}=12$~MeV with $N_{\rm bin}=6$ bins per partial wave for $\ell\leq5$. The ${}^{10}\mathrm{Be}$--${}^{64}\mathrm{Zn}$ interaction is taken from the optical-model fit of Di Pietro \emph{et al.}~\cite{DiPietro2012} and the $n$--${}^{64}\mathrm{Zn}$ interaction from KD02~\cite{KD02}.

\begin{figure}[t]
\centering
\includegraphics[width=\columnwidth]{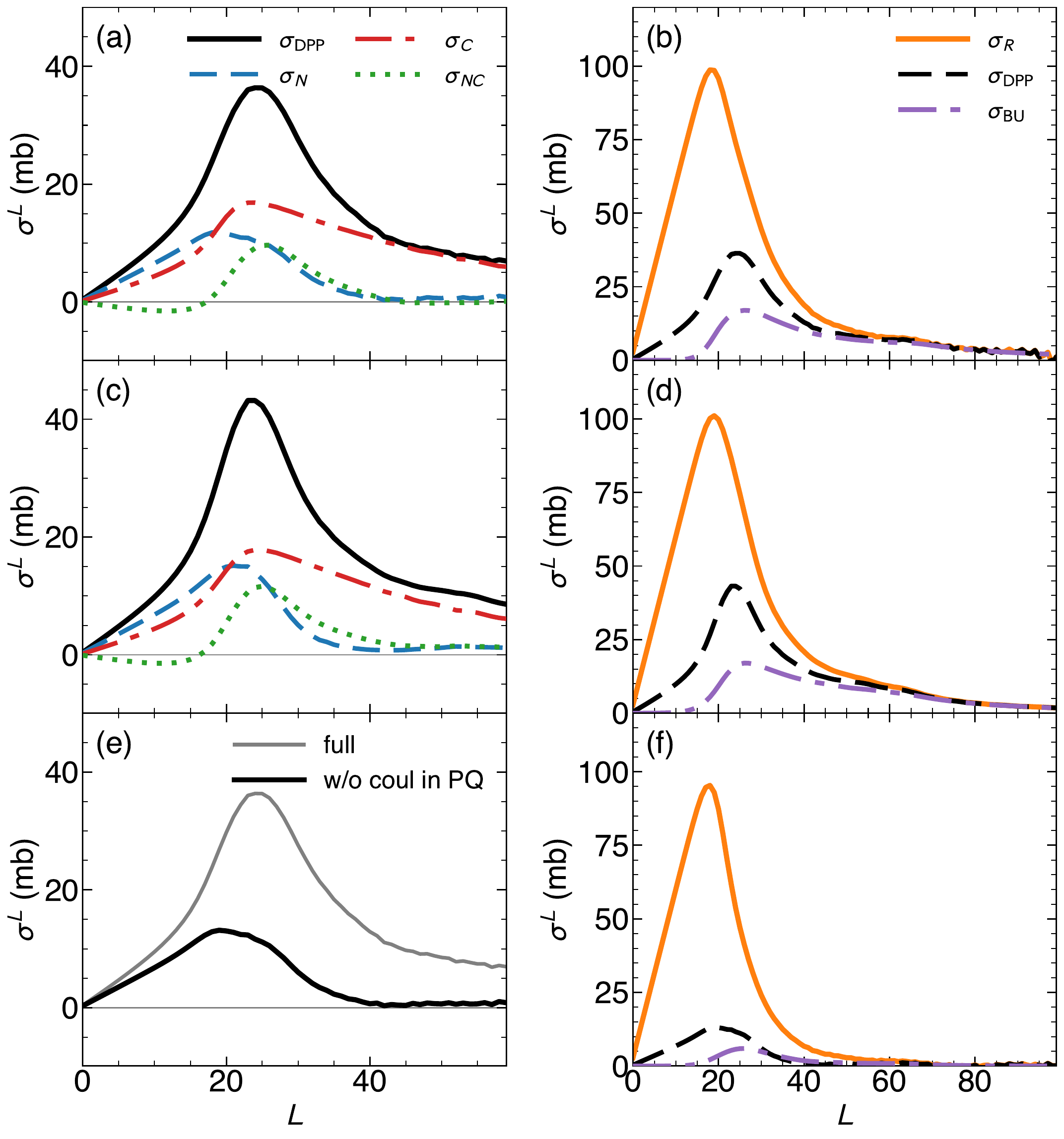}
\caption{ Partial-wave analysis for ${}^{11}\mathrm{Be}+{}^{64}\mathrm{Zn}$ at $E_{\rm lab}=28.7$ MeV. Panels (a) and (b) show the full CDCC result. Panels (c) and (d) show a diagnostic calculation in which only the off-diagonal Coulomb couplings inside the $Q$ space are removed, while the $P\!\leftrightarrow\!Q$ Coulomb bridge is retained. Panels (e) and (f) show the complementary diagnostic in which the Coulomb part of the $P\!\leftrightarrow\!Q$ bridge is removed. The left column displays the DPP decomposition, while the right column compares $\sigma_R^L$, $\sigma_{\rm DPP}^L$, and $\sigma_{\rm BU}^L$.}
\label{fig:BeZn}
\end{figure}

We now apply the decomposition to the halo case, where the Coulomb contribution is expected to be more visible. The first two systems establish the baseline and the mixed regime; Fig.~\ref{fig:BeZn} asks whether the same decomposition can identify the matrix elements that carry peripheral polarization in a weakly bound halo system, where the extended projectile density makes long-range Coulomb couplings especially effective. We first identify the dominant component in the DPP decomposition, then compare the DPP-induced absorption with the reaction and breakup profiles, and finally use two diagnostic calculations to separate the role of the $P\!\leftrightarrow\!Q$ Coulomb couplings from Coulomb propagation inside $Q$.

In the full calculation, Fig.~\ref{fig:BeZn}(a), the Coulomb contribution $\sigma_C^L$ dominates the nuclear one over the entire range of partial waves. It sets the height of the peak near $L\simeq25$ and remains the only sizeable component throughout the peripheral region $L\gtrsim35$, in sharp contrast to both previous systems. The large $B(E1)$ strength connecting the ${}^{11}\mathrm{Be}$ ground state to its low-lying dipole continuum~\cite{DiPietro2012}, together with the small neutron separation energy $S_n=504$~keV, makes the Coulomb bridge the leading entrance and exit channel over a broad range of $L$. The interference $\sigma_{NC}^L$ is strongly structured, with a small destructive dip near $L\simeq18$ followed by a constructive bump around $L\simeq25$, a signature of intricate nuclear--Coulomb phase relations in the halo wave function that have no counterpart in the systems considered above.

The observable connection is shown in Fig.~\ref{fig:BeZn}(b). At small $L$, $\sigma_R^L$ is dominated by the bare-potential absorption through $U_{00}$ and greatly exceeds both $\sigma_{\rm DPP}^L$ and $\sigma_{\rm BU}^L$, reflecting compact collisions where the ${}^{10}\mathrm{Be}$ core is absorbed by the target. At peripheral partial waves, $L\gtrsim35$, the bare-potential absorption has become negligible, so $\sigma_R^L\simeq\sigma_{\rm DPP}^L$. The generalized optical theorem within CDCC~\cite{Liu2026Coherent} further partitions $\sigma_{\rm DPP}^L$ into the asymptotic elastic-breakup yield $\sigma_{\rm BU}^L$ and a continuum-induced absorption contribution carried by the imaginary parts of the fragment--target potentials acting between breakup-bin wave functions. The latter contribution requires overlap of the bin densities with the absorptive nuclear interior and is therefore strongly suppressed at peripheral $L$, as directly visible in Fig.~\ref{fig:BeZn}(b), where the $\sigma_{\rm DPP}^L-\sigma_{\rm BU}^L$ gap closes for $L\gtrsim35$. At peripheral partial waves, the bare-potential absorption is weak and the continuum-induced absorption is also suppressed. As a result, $\sigma_R^L \simeq \sigma_{\rm DPP}^L \simeq \sigma_{\rm BU}^L$. The high-$L$ breakup yield therefore traces the same DPP contribution that dominates the reaction cross section in this region. Because this high-$L$ DPP contribution is dominated by $\sigma_C^L$, it provides an observable signal that the peripheral polarization is mainly generated by the Coulomb $P\!\leftrightarrow\!Q$ couplings, $U^{(C)}_{0\gamma}$ and $U^{(C)}_{\gamma'0}$.

The remaining panels test whether this conclusion originates from the $P\!\leftrightarrow\!Q$ Coulomb couplings or from Coulomb propagation inside $Q$. In panels (c) and (d), the off-diagonal Coulomb coupling between bin channels, $U^{(C)}_{\gamma\gamma'}$ with $\gamma,\gamma'\in Q$ and $\gamma\neq\gamma'$, is set to zero before the coupled-channel equations are solved, while all nuclear couplings, the ground-state-to-continuum Coulomb bridge, and the $Q$-$Q$ diagonal Coulomb are retained. The broad peak of $\sigma_{\rm DPP}^L$ around $L\simeq25$, the peripheral Coulomb dominance, and the high-$L$ proximity $\sigma_R^L\simeq\sigma_{\rm DPP}^L\simeq\sigma_{\rm BU}^L$ all survive. The removal of $Q$-$Q$ off-diagonal Coulomb does modify the magnitude, producing a modest enhancement of both elastic breakup and DPP-induced absorption, but it does not remove the Coulomb-bridge pattern.

The complementary diagnostic in panels (e) and (f) removes the Coulomb part of the $P\!\leftrightarrow\!Q$ bridge, $U^{(C)}_{0\gamma}$ and $U^{(C)}_{\gamma'0}$, while retaining every other coupling, including the $Q$-$Q$ Coulomb sector. This time both the DPP-induced absorption and the elastic breakup collapse to a small fraction of their full values. The residual $\sigma_{\rm DPP}^L$ is purely nuclear, remains low and narrow around $L\simeq22$, and disappears almost completely beyond $L\simeq40$. The peripheral proximity $\sigma_R^L\simeq\sigma_{\rm DPP}^L\simeq\sigma_{\rm BU}^L$ is destroyed. The contrast between panels (c,d) and (e,f) provides the evidence: the high-$L$ polarization signal of the halo system is controlled by the Coulomb bridge couplings themselves, not merely by Coulomb dynamics somewhere inside the coupled-channel Hamiltonian.


\subsection{Proton-halo extension: ${}^{8}\mathrm{B}+{}^{64}\mathrm{Zn}$}

For ${}^{8}\mathrm{B}+{}^{64}\mathrm{Zn}$ at $E_{\rm lab}=36.11$~MeV, ${}^{8}\mathrm{B}$ is described as a $p+{}^{7}\mathrm{Be}$ cluster. The $p$--${}^{7}\mathrm{Be}$ binding potential is taken from Esbensen and Bertsch~\cite{Esbensen1996}; since spin degrees of freedom are neglected, the well depth is rescaled to reproduce the proton separation energy in a spinless configuration, as is standard in spinless CDCC treatments of weakly bound projectiles. The continuum is discretized up to $\varepsilon_{\max}=12$~MeV with $N_{\rm bin}=6$ bins per partial wave for $\ell\leq3$. The $p$--${}^{64}\mathrm{Zn}$ optical potential is taken from the KD02 global parametrization~\cite{KD02}, and the ${}^{7}\mathrm{Be}$--${}^{64}\mathrm{Zn}$ potential from the ${}^{7}\mathrm{Li}$ elastic-scattering fit of Cutler \emph{et al.}~\cite{Cutler1977}.

After establishing the Coulomb-bridge behavior in ${}^{11}\mathrm{Be}+{}^{64}\mathrm{Zn}$, we use ${}^{8}\mathrm{B}+{}^{64}\mathrm{Zn}$ as a further test in a proton-halo system. Two features amplify the Coulomb bridge for a proton halo. First, the smaller proton separation energy, $S_p({}^{8}\mathrm{B})=137$~keV compared with $S_n({}^{11}\mathrm{Be})=504$~keV, makes the proton-halo wave function more extended than the neutron-halo one, so that the $P\!\leftrightarrow\!Q$ form factors built on the halo density reach further into the long-range Coulomb region. 
Second, the halo nucleon in ${}^{8}\mathrm{B}$ is itself charged. The proton--target Coulomb interaction can therefore enter the Coulomb transition matrix element $U^{(C)}_{0\gamma}$ directly, rather than only through the motion of a charged core relative to a neutral valence particle. This also connects naturally with the post-acceleration discussed in Ref.~\cite{DIPIETRO2019134954}. After breakup, the charged fragments continue to be separated and accelerated by the target Coulomb field. In the present decomposition, this physics appears as an enhanced Coulomb sector of the $P\!\leftrightarrow\!Q$ coupling and of the subsequent breakup dynamics. Although the leading dipole effective charges of the two halos are comparable in magnitude ($\simeq 0.36\,e$ for ${}^{11}\mathrm{Be}$ and $\simeq 0.38\,e$ for ${}^{8}\mathrm{B}$, with opposite sign), the proton-halo case contains an additional direct proton--target Coulomb contribution. The question is therefore whether the same peripheral relation, $\sigma_{\rm BU}^L\simeq\sigma_{\rm DPP}^L\simeq\sigma_C^L$, also appears in this case, and how the charged valence proton affects the nuclear--Coulomb interference.

\begin{figure}[t]
\centering
\includegraphics[width=\columnwidth]{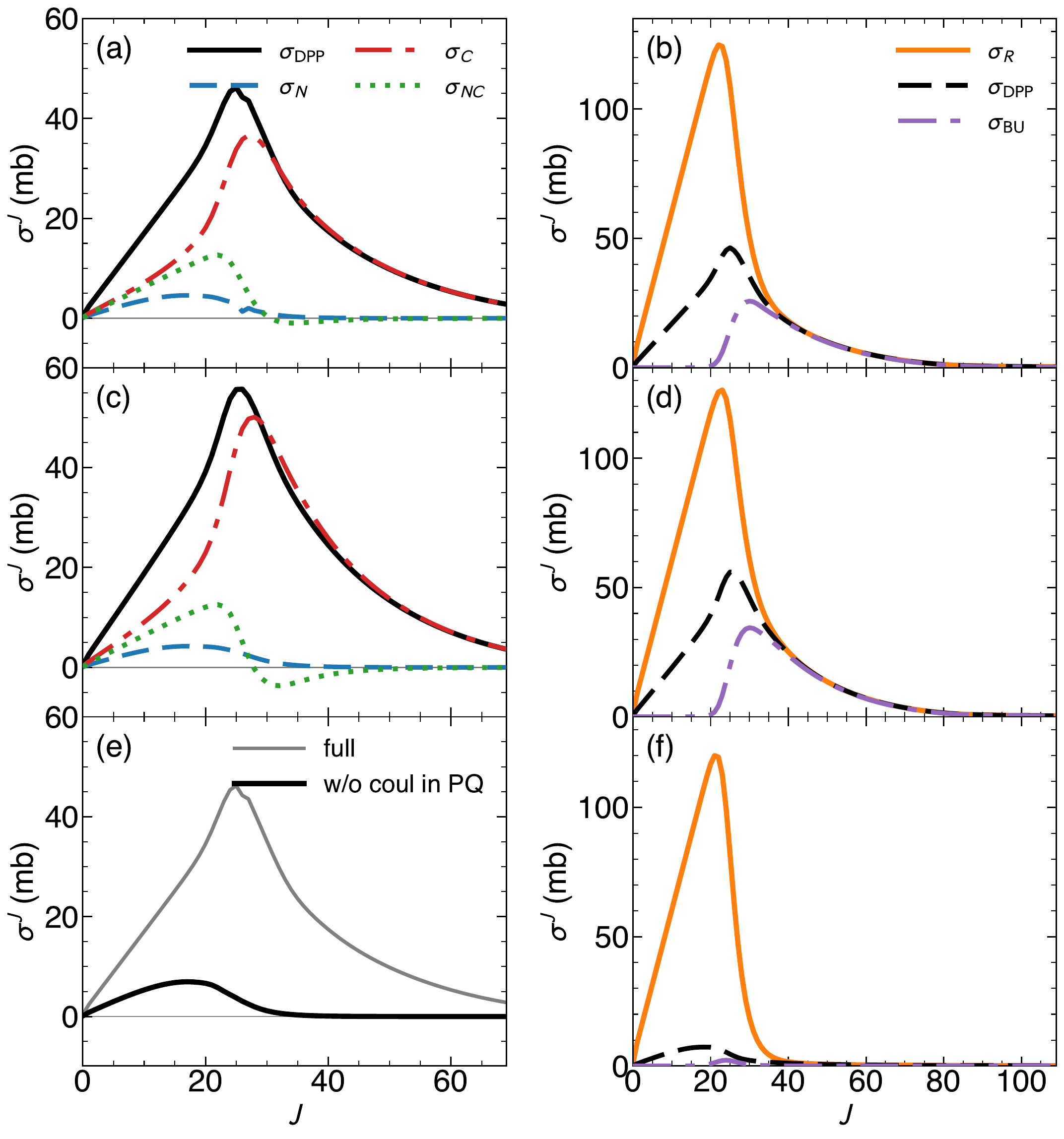}
\caption{Partial-wave analysis for ${}^{8}\mathrm{B}+{}^{64}\mathrm{Zn}$ at $E_{\rm lab}=36.11$ MeV. Panels (a) and (b) show the full CDCC result. Panels (c) and (d) show a diagnostic calculation in which the off-diagonal Coulomb couplings inside the $Q$ space are removed, while the direct $P\!\leftrightarrow\!Q$ Coulomb coupling is retained. Panels (e) and (f) show the complementary diagnostic in which the Coulomb part of the direct $P\!\leftrightarrow\!Q$ coupling is removed, while the $Q$-$Q$ Coulomb sector is retained. The left column displays the DPP decomposition, while the right column compares $\sigma_R^L$, $\sigma_{\rm DPP}^L$, and $\sigma_{\rm BU}^L$.}
\label{fig:BZn}
\end{figure}

The full calculation in Figs.~\ref{fig:BZn}(a) and \ref{fig:BZn}(b) shows that this relation is indeed realized. The Coulomb component $\sigma_C^L$ closely follows the total $\sigma_{\rm DPP}^L$ over almost the whole range of partial waves and determines the main peak near $L\simeq25$, while the nuclear component remains a small residual. At larger partial waves, where the bare absorption through $U_{00}$ becomes weak, the reaction, DPP-induced absorption, and elastic-breakup profiles approach one another, $\sigma_R^L\simeq\sigma_{\rm DPP}^L\simeq\sigma_{\rm BU}^L$. Thus the high-$L$ breakup yield follows a DPP contribution that is almost entirely Coulomb dominated. Compared with ${}^{11}\mathrm{Be}+{}^{64}\mathrm{Zn}$, the Coulomb dominance is stronger, and the interference term $\sigma_{NC}^L$ becomes positive in the main peak region, indicating constructive nuclear--Coulomb interference in this proton-halo case.

The two diagnostic calculations confirm that this behavior is controlled by the direct $P\!\leftrightarrow\!Q$ Coulomb coupling. When the off-diagonal Coulomb couplings inside the $Q$ space are removed, Figs.~\ref{fig:BZn}(c) and \ref{fig:BZn}(d), the main features of the full calculation remain. The DPP is still dominated by $\sigma_C^L$, and the high-$L$ relation $\sigma_R^L\simeq\sigma_{\rm DPP}^L\simeq\sigma_{\rm BU}^L$ is preserved. By contrast, when the Coulomb part of the direct $P\!\leftrightarrow\!Q$ coupling is removed, Figs.~\ref{fig:BZn}(e) and \ref{fig:BZn}(f), both $\sigma_{\rm DPP}^L$ and $\sigma_{\rm BU}^L$ are strongly suppressed and the high-$L$ relation disappears. The ${}^{8}\mathrm{B}$ calculation therefore supports the same conclusion reached for ${}^{11}\mathrm{Be}$, namely that peripheral breakup affects the elastic channel mainly through the Coulomb part of the direct $P\!\leftrightarrow\!Q$ coupling. In the proton-halo case this mechanism is further enhanced by the charge of the valence proton.

\section{Discussion}
\label{sec:discussion}

Our results refine the qualitative picture developed by Di Pietro \emph{et al.}~\cite{DiPietro2010,DiPietro2012} for ${}^{11}\mathrm{Be}+{}^{64}\mathrm{Zn}$, and by Spart\`a \emph{et al.}~\cite{Sparta2021} for ${}^{8}\mathrm{B}+{}^{64}\mathrm{Zn}$. Those authors argued, by switching off the high-multipole Coulomb and high-multipole nuclear couplings while retaining only the monopole diagonal terms, that the Coulomb interaction is essential to reproduce the suppression of the Coulomb--nuclear interference peak in the elastic angular distribution. The switch-off analysis tells us that the Coulomb sector matters; our operator-level decomposition addresses a different question and asks which $P\!\leftrightarrow\!Q$ bridge couplings carry the peripheral polarization once a common propagator is fixed. 

The decomposition~(\ref{eq:dpp_components}) identifies the physical conduit responsible. It is not the mere presence of Coulomb strength in the Hamiltonian but the Coulomb sector of the $P\!\leftrightarrow\!Q$ bridge couplings $U_{0\gamma}$ and $U_{\gamma'0}$, the link that connects the elastic channel to the reaction space, that governs the polarization at peripheral angular momenta. In light-to-intermediate systems the nuclear bridge is strong enough that this distinction is not crucial for the integrated absorption, and an analysis that ignores the Coulomb coupling would capture the essential physics. In the halo configuration, however, the Coulomb side of the bridge becomes the dominant mechanism: the high-$L$ structure of $\sigma_{\rm DPP}^L$, together with the near coincidence of $\sigma_R^L$, $\sigma_{\rm DPP}^L$, and $\sigma_{\rm BU}^L$ in ${}^{11}\mathrm{Be}+{}^{64}\mathrm{Zn}$ and in ${}^{8}\mathrm{B}+{}^{64}\mathrm{Zn}$, shows that the peripheral polarization is fed mainly through $U^{(C)}_{0\gamma}$ and $U^{(C)}_{\gamma'0}$. In that sense the high-$L$ breakup cross section is supported by the same bridge; it exposes the mechanism observationally, but the mechanism itself resides in the bridge.


The nuclear--Coulomb interference reflects how the nuclear and Coulomb parts of the $P\!\leftrightarrow\!Q$ coupling combine with each other. Its sign depends on the relative phase between the short-range nuclear form factor and the long-range Coulomb form factor. For $d+{}^{58}\mathrm{Ni}$ and ${}^{6}\mathrm{Li}+{}^{208}\mathrm{Pb}$, this interference is mainly negative and partly cancels the Coulomb contribution. For ${}^{11}\mathrm{Be}+{}^{64}\mathrm{Zn}$, the interference changes with $L$ but remains weakly negative in low $L$. For ${}^{8}\mathrm{B}+{}^{64}\mathrm{Zn}$, it becomes constructive near the main peak. This difference can be understood from the Coulomb coupling itself. In the neutron-halo case ${}^{11}\mathrm{Be}$, the valence neutron is neutral, so the Coulomb effect mainly comes from the charged core interacting with the target. This is related to the post-acceleration dynamics discussed in Ref.~\cite{DIPIETRO2019134954}. In the proton-halo case ${}^{8}\mathrm{B}$, the valence proton is charged and therefore gives an additional direct proton--target Coulomb contribution to $U^{(C)}_{0\gamma}$. This additional contribution changes the relative phase between the nuclear and Coulomb form factors and makes the interference constructive near the peak.

In every case considered, direct sums $\sigma_N+\sigma_C$ misrepresent the DPP absorption. The sign and size of the discrepancy track the system: $d+{}^{58}\mathrm{Ni}$ shows only a small net residue after $\sigma_C$ and $\sigma_{NC}$ partially cancel; ${}^{6}\mathrm{Li}+{}^{208}\mathrm{Pb}$ shows a strongly destructive nuclear--Coulomb interference that pulls $\sigma_{\rm DPP}$ well below $\sigma_N+\sigma_C$; the neutron-halo ${}^{11}\mathrm{Be}+{}^{64}\mathrm{Zn}$ case shows an $L$-structured $\sigma_{NC}^L$ whose integrated value remains weakly destructive; the proton-halo ${}^{8}\mathrm{B}+{}^{64}\mathrm{Zn}$ case has the opposite sign, with constructive $\sigma_{NC}$ that would make $\sigma_N+\sigma_C$ underestimate $\sigma_{\rm DPP}$ near the peak. A consistent treatment of $\Delta U_{NC}$ at the operator level is therefore essential whenever the Coulomb and nuclear sectors of the bridge are comparable.

\section{Summary}
\label{sec:summary}

We have analyzed how continuum excitation feeds back into elastic scattering by decomposing the bridge couplings of the Feshbach dynamical polarization potential. Within CDCC, the splitting of the $P\!\leftrightarrow\!Q$ couplings into nuclear and Coulomb parts gives $\Delta U_{\rm DPP}=\Delta U_N+\Delta U_C+\Delta U_{NC}$, with all three components sharing the same $Q$-space Green's function. The same separation applies partial wave by partial wave to the DPP-induced absorption, making it possible to identify which bridge carries polarization flux out of the elastic channel.

The four reactions studied here form a hierarchy of that bridge. In $d+{}^{58}\mathrm{Ni}$ the DPP absorption is governed by the nuclear bridge, with the Coulomb component largely cancelled by interference. In ${}^{6}\mathrm{Li}+{}^{208}\mathrm{Pb}$ the Coulomb bridge becomes peripheral and competitive, but it remains part of a strongly interfering nuclear--Coulomb mixture. In ${}^{11}\mathrm{Be}+{}^{64}\mathrm{Zn}$, the halo limit is reached: for $L\gtrsim35$, the reaction cross section is driven mainly by the DPP, the DPP is dominated by the Coulomb bridge, and the corresponding breakup yield is supported by the same high-$L$ sector. The proton-halo case ${}^{8}\mathrm{B}+{}^{64}\mathrm{Zn}$ amplifies the same mechanism, with the Coulomb bridge accounting for essentially the entire $\sigma_{\rm DPP}^L$ and the nuclear--Coulomb interference now adding constructively. The present analysis therefore turns the qualitative statement that Coulomb couplings matter into a microscopic statement about where the peripheral polarization is carried, namely by $U^{(C)}_{0\gamma}$ and $U^{(C)}_{\gamma'0}$. Extensions to refractive observables sensitive to $\operatorname{Re}\Delta U_{\rm DPP}$ and to reactions on other halo nuclei such as ${}^{6}\mathrm{He}$ are in progress.

\section*{Acknowledgements}

We thank Antonio M.\ Moro for useful discussions. This work was supported by the National Natural Science Foundation of China (Grant Nos.\ 12535009 and 12475132), the National Key R\&D Program of China (Contract No.\ 2023YFA1606503), and the Fundamental Research Funds for the Central Universities.

\bibliography{ref}

\end{document}